\documentstyle[11pt,aaspp4]{article}

\def\temperature{10^8 \K}
\def\density{10^{-3}  \percc }
\def\mass{2 \times 10^{15} \, h_{50}^{-1} \MSun}
\def\equilibrium{t_{\rm eq}}
\def\Tx{T_{\rm x}}
\def\Heion{ {\rm He}^{++}}
\def\sthreehalves{{\textstyle {3 \over 2}}}
\def\sound{c_S}
\def\turbulent{v_{\rm turb}}
\def\los{v_{los}}

\def\xiv{\hbox to .015em{$\xi$\hss}\hbox to .015em{$\xi$\hss}\xi}
\def\omvec{\hbox to .015em{$\omega$\hss}\hbox to .015em{$\omega$\hss}\omega}

\def\muvec{\hbox to .015em{$\mu$\hss}\hbox to .015em{$\mu$\hss}\mu}
\def\betavec{\hbox to .015em{$\beta$\hss}\hbox to .015em{$\beta$\hss}\beta}

\def\keV{\, {\rm keV}}
\def\eV{\, {\rm eV}}
\def\K{\, {\rm K}}

\def\yr{\, {\rm yr}}

\def\Mpc{\, {\rm Mpc}}
\def\percc{\, {\rm cm}^{-3}}
\def\microgauss{\, \mu {\rm G}}

\def\prop{\propto}

\def\defeq{\equiv}

\def\Xeotn/{${}^{129}\rm{Xe}$}

\def\half{{1 \over 2}}
\def\shalf{{\textstyle {1 \over 2}}}
\def\sthreehalves{{\textstyle {3 \over 2}}}

\def\MSun{M_{\odot}}

\begin{document}

\title{Do the Electrons and Ions in X-ray Clusters Share the Same Temperature?}
\author{David C. Fox\altaffilmark{1}
\altaffiltext{1}{Also at: Physics Department, Harvard University}
and Abraham Loeb}
\affil{Astronomy Department, Harvard University,\\ 
60 Garden St., Cambridge MA 02138}

\begin{abstract}
The virialization shock around an X-ray cluster primarily heats the ions,
since they carry most of the kinetic energy of the infalling gas.
Subsequently, the ions share their thermal energy with the electrons
through Coulomb collisions. We quantify the expected temperature difference
between the electrons and ions as a function of radius and time, based on a
spherical self-similar model for the accretion of gas by a cluster in an
$\Omega=1$, $h_{50}=1$ universe. Clusters with X-ray temperatures
$\Tx=4$--$10\times 10^7 \K$, show noticeable differences between their
electron and ion temperatures at radii $\ga 2$Mpc.  High resolution
spectroscopy with future X-ray satellites such as Astro~E may be able to
determine the ion temperature in intracluster gas from the width of its
X-ray emission lines, and compare it to the electron temperature as
inferred from the free-free emission spectrum.  Any difference between
these temperatures can be used to date the period of time that has passed
since the infalling gas joined the cluster.

\end{abstract}

\keywords{galaxies: clusters of -- cosmology: theory}

\section{Introduction}

The standard technique for measuring the total mass in clusters of galaxies
from X-ray observations assumes that the cluster gas is in hydrostatic and
thermal equilibrium (see, e.g.  Jones \& Forman 1992).  Hydrodynamic
simulations often maintain the assumption of thermal equilibrium, and show
that while the gas is typically at hydrostatic equilibrium in the cores of
clusters, its state is perturbed in their outer parts (Navarro, Frenk, \&
White 1995; Evrard, Metzler, \& Navarro 1996).  In this paper we address
the question of whether the gas in these outer regions should actually be
in thermal equilibrium as assumed.  To see why the cluster gas might deviate
from thermal equilibrium we should first examine how the gas obtains its
high X-ray temperature.

As fresh material joins the stationary cluster gas, it converts its infall
kinetic energy into heat through a shock. Since most of the inertia of the
infalling gas is carried by the ions, they acquire most of the dissipated
energy and heat to about twice the virial temperature of the cluster.  The
electrons, on the other hand, remain cold after the shock with their
temperature slowly rising as a result of their Coulomb collisions with the
hot ions. Thermal equilibrium will only be achieved if the temperature
equilibration time between the ions and the electrons is shorter than the
age of the cluster. In fact, the ratio between the equilibration time and
the age of the Universe is of order unity, $\sim (\Omega_b h_{50}/0.06)
(T/10^8~{\rm K})^{3/2} (\delta/178)^{-1}$, for the characteristic
overdensity $\delta$ and temperature $T$ in cluster shocks, where 
$\Omega_b$ is the baryonic density parameter and $h_{50}$ is the Hubble
constant in units of $50~{\rm km~s^{-1}~Mpc^{-1}}$.  This implies that the
electron and ion fluids might not achieve complete thermal equilibrium
within the cluster lifetime.

A difference between the electron and ion temperatures was invoked as a
potential explanation for the increase in the inferred baryon fraction in
the outer parts of A2163 (Markevitch et al. 1996).  The ASCA data show a
sharp drop in the electron temperature beyond $2 \, h_{50}^{-1} \, {\rm
Mpc}$ in this cluster.  The gas distribution, determined by ROSAT, is well
described by a $\beta$-model.  Assuming hydrostatic equilibrium, a dark
matter distribution with the same profile as the gas was strongly ruled out
by the temperature measurements, and models in which the dark matter fell
off much more rapidly were only marginally allowed.  If, however, the
electron temperature is lower than the ion temperature, the actual gas
pressure is higher than the value inferred under the assumption of thermal
equilibrium.  Markevitch et al.  (1996) point out that this possibility
might be real since the equilibration timescale is comparable to the
typical time since the last merger.

In principle, the deviation from thermal equilibrium could result either
from smooth accretion or from mergers of clumps. For simplicity, we focus on
smooth accretion in this paper. Making use of the versatility of a
semi-analytic approach, we employ a self-similar model of spherical
accretion to examine the regime of validity of the assumption of thermal
equilibrium between the electrons and ions.  This approach is similar to
that of Itoh (1978), who considered thermal equilibration in supernova
remnants using a self-similar blast-wave model.  We assume that Coulomb
collisions are the only coupling mechanism between the temperatures of
electrons and ions.  It can be shown that a homogeneous plasma with
isotropic, monotonically decreasing distribution functions is stable
against plasma instabilities (Rosenbluth 1965).  Additional coupling might
be provided by plasma instabilities in more complicated conditions, but we
ignore them in this discussion.

In \S 2 we describe the method of the calculation. In \S 3 we relate the
parameters of our model to the conditions in typical clusters.  In \S 4
we describe the details of the numerical calculations.  In \S 5 we
quantify the effects of the approximations made in our model.
Finally, \S 6 discusses the results and the possibility of detecting
the difference
between the ion and electron temperatures with future X-ray telescopes.
Throughout the paper we assume a density parameter $\Omega=1$, a Hubble
constant $h_{50}=1$, and a hydrogen mass fraction $X=76 \%$.

\section{Method of Calculation}
\subsection{Collision Timescales}

We first consider collisional relaxation processes in plasmas consisting of
several species of particles.  Spitzer (1962) showed that each species
achieves a Maxwellian distribution on a timescale of order the collision
time,
\begin{equation}
t_{xx} = {m_x^{ 1 / 2} (3kT)^{3/2} \over 5.71 \pi n e^4 Z^4 \ln \Lambda}
= 3.60 \times 10^8 \yr  \, {(T / \temperature)^{3/2} \over
(n / \density)}
 {A^{ 1 / 2} \over Z^4 \ln \Lambda},
\end{equation}
where $A = m_x/m_p$ is the particle mass of the species $x$ in units of the
proton mass, $Ze$ is the particle charge, and $\ln \Lambda$ is the Coulomb
logarithm.  For temperatures above $4
\times 10^5 \K$,
\begin{equation}
\ln \Lambda \approx 37.8 + \ln (T / \temperature) -\half \ln (n / \density).
\end{equation}
Thus, the collision time for protons, $t_{pp} =(m_p/m_e)^{1/2} \, t_{ee}$, is
longer by a factor of 43 than that for electrons, $t_{ee}$.

Spitzer also found that the exchange of energy in collisions between test
particles and field particles, with Maxwellian distributions of
temperatures $T$ and $T_f$, is governed by the equation
\begin{equation}
{d T \over  dt} = {{T_f-T} \over \equilibrium},
\end{equation}
where the equilibration timescale, $\equilibrium$, depends on the related
particle masses, $m$ and $m_f$, and charge numbers, $Z$ and $Z_f$, as
well as the density of field particles, $n_f$, and is given by
\begin{eqnarray}
\equilibrium &=& {3m m_f \over 8 (2 \pi)^{ 1 / 2} n_f Z^2 Z_f^2 e^4 \ln
\Lambda} \left ({k T \over m} + {k T_f \over m_f} \right)^{3/2}
\nonumber \\
 & = & {1.86 \times 10^8\yr}  {A_f \over Z^2 Z_f^2 A^{1/2} \ln \Lambda}
{(T/ \temperature)^{3/2} \over (n_f/ \density)}
\left (1 + {T_f \over T} {A \over A_f} \right)^{3/2}. 
\end{eqnarray}
For collisions between protons and ions of similar mass, such as $\Heion$, the
equilibration timescale is comparable to $t_{pp}$.  For electrons and
ions, the equilibration timescale is greater than $ t_{pp}$ by a factor of
order $(m_p/m_e)^{1/2}$.

Thus, equilibrium in an electron-proton plasma is achieved in several
stages.  First, the electrons reach a Maxwellian distribution with
temperature $T_e$ on a timescale $t_{ee}$.  Then, on a longer timescale,
$t_{pp} \sim (m_p/m_e)^{ 1 / 2} t_{ee}$, the protons reach a Maxwellian
distribution with temperature $T_p$.  Finally, the two temperatures
equalize on a timescale of order $t_{ep} \sim (m_p/m_e) t_{ee}$.
We would like to consider a fully ionized plasma of hydrogen and helium,
which is only slightly more complicated.  All three ion timescales,
$t_{\Heion p}$, $t_{pp}$, and $t_{\Heion\Heion}$, are much shorter than the
electron-ion equilibration timescale.  Therefore, when considering
electron-ion equilibration, we can regard the ions as sharing a single
temperature $T_i$.  In this case,
\begin{equation}
{d T_e \over dt} = {T_i-T_e \over t_{ei}}
\end{equation}
where the energy exchange rate between the ions and the electrons is just
the sum of the rates due to exchange with the two ion species,
\begin{equation}
1/t_{ei} = 1/t_{ep} + 1/t_{e\Heion}.
\end{equation}

We will generally be in the regime in which $T_e/T_i \gg m_e/m_p$, where
the contribution from each species to $t_{ei}^{-1}$ is proportional to $n_f
Z_f^2/A_f$.  For both $p$ and $\Heion$, this factor simply equals the
field particle density, $n_f$.  Thus, in the case of a fully ionized plasma of
hydrogen and helium,
\begin{equation}
t_{ei} = {7.97 \times 10^9 \yr} {(T_e
/ \temperature)^{3/2} \over (n_i / \density)\ln \Lambda}
\end{equation}
where $n_i$ is the total ion number density.  This timescale
differs from  $t_{ep}$ only by a factor of $n_p/n_i$.

\subsection{Self-Similar Model for Spherical Accretion}

For simplicity, we consider spherical accretion of gas onto an X-ray
cluster in an $\Omega = 1$ universe.  Bertschinger (1985) found that an
initial top-hat mass perturbation results in a self-similar solution, 
for both purely collisional and purely collisionless fluids, as well as for a single
collisional fluid in a potential well dominated by a collisionless fluid
(i.e. $\Omega_b \ll \Omega = 1$.)

We would like to adapt this last model to the case of a fully ionized
plasma of hydrogen and helium, under the gravitational influence of
collisionless dark matter.  The gas in Bertschinger's model is considered
adiabatic, and we retain this simplifying assumption.  Thus, we neglect
radiative cooling, a well-justified assumption for the outer parts of
clusters where the free-free cooling time is much longer than the Hubble
time.  We also assume that there exists a magnetic field which is
sufficiently strong to suppress thermal conduction, yet small enough to
have a negligible dynamical effect.  These conditions are indeed satisfied
by the inferred magnetic field strength $\sim 0.1-1 \microgauss$ in cluster
environments (Kim et al. 1990; Kim, Tribble, \& Kronberg 1991; Rephaeli
1988).
The general case of a stationary shock in a fully ionized plasma with
conduction is complicated by the fact that the electron thermal speed
exceeds the shock speed, allowing the electrons to preheat the plasma ahead
of the shock (Shafranov 1957; Zel'dovich \& Raizer 1967).  If conduction is
suppressed by a magnetic field, however, we will see that electron-ion
equilibrium can be treated entirely locally, which simplifies the problem
considerably.

The self-similar model is divided into three regions: pre-shock, shock, and
post-shock.  In the pre-shock region, the gas is taken to be cold, so that
pressure is neglected.  In our model, we take the plasma to be fully
ionized even before the shock, so it cannot literally have zero temperature
(e.g., photoionization by the UV background would heat the gas to $\sim
10^{4-5} \K$).  However, as long as the temperature is sufficiently low
that the initial pressure is dynamically insignificant compared to gravity
(i.e.  $T\ll 10^7 \K$ for X-ray clusters), this will not affect the
pre-shock behavior. In this case, the acceleration is due to gravity alone,
and thus the pre-shock boundary conditions for the gas density, pressure,
and velocity ($\rho$, $p$, and $v$) are the same as in the case of a
single, cold, neutral gas.

The behavior of the gas at the shock is determined by the shock jump
conditions for $\rho$, $p$, and $v$.  These conditions derive from the
conservation of total mass, momentum, and energy.  Since we neglect
conduction, the equations relating the pre-shock density, pressure, and
velocity, $\rho_1$, $p_1$, and $v_1$, to the post-shock values, $\rho_2$,
$p_2$, and $v_2$, are the same for an ionized plasma as for a single
monoatomic gas.  Since the pre-shock flow is cold, $p_1$ is negligible
compared to $p_2$, and the shock is strong.

Even though the relation between pre-shock and post-shock conditions is
unchanged, the self-similar solution could still be altered by the finite
shock thickness.  The thickness of collisional shocks is typically of order
the mean free path, which, for Coulomb collisions, is proportional to
$T^2$.  For the ions, whose post-shock temperature is high, the mean free
path in the post-shock region may be large.  In the hottest clusters, at $z
= 0$, it can become a non-negligible fraction of the shock radius, $r_s$,
which characterizes the scale of the solution.  This could break the
self-similarity of the solution.  In the presence of the magnetic field
mentioned above, however, a collisionless shock, which can be much thinner
than a mean free path, is more likely.  We will neglect the finite
thickness of the shock in our model.

The shock jump conditions determine the mean temperature, $\overline T$,
given the mass per particle, $\mu m_p$.  However, in order to find $T_i$
and $T_e$ separately, we must consider the shock region in more detail.
Inside the shock, collisions convert bulk kinetic energy into thermal
energy.  The increase, $\sthreehalves k \Delta T$, in thermal energy per
particle is of order $m(v_1-v_2)^2$.  Because of the ratio of masses,
$\Delta T$ is clearly much greater for the ions than for the electrons.
Adiabatic compression increases the electron temperature by a factor of $
(\rho_2/\rho_1)^{\gamma -1}$, where $\gamma$ is the ratio of specific
heats.  However, $\rho_2/\rho_1 = (\gamma + 1)/(\gamma -1)$ for a strong
shock, so for $\gamma = 5/3$, the electron temperature increases only by a
factor of $2.5$.  Thus, as long as the pre-shock temperature, $T_1$, is
negligible compared to the post-shock temperature, $T_2$, adiabatic
compression will not produce a value of $T_{e2}$ which is significant
compared to $T_{i2}$.  There could also be collisional energy exchange
within the shock.  As long as the shock is thin, the time spent in the
shock will be much less than $t_{ei}( T_e =
\overline T)$.  When $T_e \ll \overline T$, $t_{ei}$ is shorter by
$(T_e/\overline T)^{3/2}$, but we will see in \S 5 that this has only a
small effect at later times.  We therefore assume that immediately after
the shock, the ions have all of the thermal energy, and that $T_{e2} = 0$.

In a thin shock, it is possible that the ion species will not have time
to relax to Maxwellian velocity distributions or to equilibrium with
each other through collisions.  However, since the contribution per ion
to $t_{ei}^{-1}$ is the same for $p$ and $\Heion$, $d T_e/dt$ depends
only on the mean ion temperature.  Furthermore, once the ions reach
equilibrium with each other, the rate of change of their temperatures
due to collisions with electrons is the same, so
they will remain in equilibrium.  Thus, for simplicity, we assume a
single ion temperature, $T_i$.

In the post-shock region, the adiabatic equation must be replaced by an
energy equation for each species, describing $d T/dt$ due to collisions as
well as adiabatic compression.  Assuming a single ion temperature
reduces the number of energy equations to two: (i) an
electron equation,
\begin{equation}
{d T_e \over dt} = {T_i-T_e \over t_{ei}} + (\gamma -1) {T_e \over n}
{dn \over dt},
\label{electron_energy}
\end{equation}
describing the temperature evolution of electrons in a given fluid element,
and (ii) an equation for the mean temperature of the particles in the same
fluid element,
\begin{equation}
{d  \overline{T} \over dt} = (\gamma -1) {\overline{T} \over n} {dn \over dt},
\end{equation}
which is obtained by averaging the equations for the individual species.

In principle, there are also separate continuity and momentum equations
for each species.  However, the Coulomb force prevents charge separation
on scales larger than the Debye length, and spherical symmetry precludes
electrical currents without charge separation, so the electron and ion
velocities are identical, and separating these equations provides no
additional information.  Furthermore, the total pressure, $p$, which
appears in the total momentum equation, depends only on the mean
temperature, $\overline{T}$, which is independent of the extent of
electron-ion equilibration and is governed by the same adiabatic
equation as in the case of a single collisional gas.  Thus, the fluid
equations for the density, pressure, and velocity are unchanged, and
are merely supplemented by  equation~(\ref{electron_energy})
which describes the temperature equilibration.  This is important, since the
equilibration timescale would otherwise break the self-similarity of
the solutions for $\rho$, $p$, and $v$.
 
Since, in the zero conduction case, the equations and boundary
conditions determining $\rho$, $p$, and $v$ in all three regions of
the self-similar model are unaffected by the replacement of a
single collisional gas with a fully ionized plasma, we obtain the exact
same solution for these variables.  Thus, we can regard Bertschinger's
solution as a background in which we can solve for $T_e$ according to
equation~(\ref{electron_energy}).

The self-similar solution has a single length scale, which we take to be
the turnaround radius, $r_{ta} (t)$.  In the case of a top-hat
perturbation, $r_{ta}\propto t^{8/9}$.  We adopt the dimensionless
functions from Bertschinger (1985) and write $\rho$, $p$, and $v$ as
functions of $\lambda
\defeq r/r_{ta}$:
\begin{eqnarray}
\rho (r, t) & = & \rho_c \Omega_b D (\lambda), 
\nonumber \\
p (r, t) & = &  \rho_c \Omega_b \left ({r_{ta} \over t} \right )^2 P
(\lambda), 
\nonumber \\
 v (r, t) & = &  {r_{ta} \over t}V (\lambda).
\label{scalings}
\end{eqnarray}
The scaling in terms of the critical density $\rho_c = (6 \pi G t^2)^{-1}$
follows from dimensional analysis, since far outside the initial top-hat
perturbation, there is no other density scale.  The factors of $\Omega_b$
are convenient in the $\Omega_b \ll 1$ case.  When we consider the
collisionless dark matter generating the potential well, we will also need
the total mass of the system
\begin{equation}
m_{\rm tot}(r,t) = {4 \pi \over 3} \rho_c r_{ta}^3 M (\lambda).
\label{mass_scaling}
\end{equation}

\subsection{Electron-Ion Equilibration}

It is useful to introduce scaled temperatures, $\widetilde T_e \defeq
T_e/\overline T$ and $\widetilde T_i \defeq T_i/\overline T$.  We then find
\begin{equation}
{d \widetilde T_e \over dt} = {1 \over \, \overline T \,} {d T_e \over dt}
-{\widetilde T_e \over \overline T^2} {d \overline T \over dt}
= {\widetilde T_i-\widetilde T_e \over t_{ei}},
\end{equation}
where the adiabatic compression terms cancel.  Note that $t_{ei}$ still
depends on $T_e= \overline T \widetilde T_e$.  However, under the
post-shock adiabatic conditions, $\overline T
\propto n_i^{\gamma -1} \propto n_i^{2/3}$, so by coincidence the factor
$\overline T^{3/2}/n_i$ is actually a constant for a given fluid element.
There is still some residual dependence in the Coulomb term, $\Lambda$, but
since $t_{ei}$ depends only logarithmically on $\Lambda$, this dependence
can be neglected.  We consider any fluid element just after the time,
$t_s$, when it passed through the shock.  It is convenient to define
$t_{2s}$ to be the value which $t_{ei}$ would have had just after the
shock if $\widetilde T_e$ were unity, i.e.
\begin{equation}
t_{2s} \defeq  t_{ei} (t_s) \widetilde T_e^{-3/2} (t_s)
 = {7.97 \times 10^9 \yr \over \ln \Lambda_{2s}}
 {(\overline T_2 (t_s)
/ \temperature)^{3/2} \over (n_{i2} (t_s) / \density)}
\end{equation}
where $\Lambda_{2s}$ is, similarly, the value $\Lambda$ would have had just
inside the shock radius, if $\widetilde T_e$ were $1$.  Then, neglecting
the slight variation in $\ln \Lambda$,

\begin{equation}
t_{ei} = t_{2s} \widetilde T_e^{3/2}
\label{timescale_scaling}
\end{equation}
and
\begin{equation}
{d \widetilde T_e \over dt} = {\widetilde T_i-\widetilde T_e \over
t_{2s} \widetilde
T_e^{3/2}}.
\label{all_scaled}
\end{equation}
Finally, by definition,
\begin{equation}
\overline T \defeq {n_e T_e + n_i T_i \over n_e + n_i},
\end{equation}
and by eliminating $\widetilde T_i$ from equation~(\ref{all_scaled}), we find
\begin{equation}
{d \widetilde T_e \over dt} = {1 \over t_{2s}} \left({n_i + n_e \over n_i}
\right) (1-\widetilde T_e) \widetilde T_e^{-3/2}.
\label{differential}
\end{equation}

For $(1-\widetilde T_e) \ll 1$, this simply represents exponential decay
of the fractional temperature difference, $\widetilde T_i-\widetilde
T_e$, with an e-folding time of $t_{2s} n_i/(n_i + n_e) \approx t_{2s}/2$.
In fact,
equation~(\ref{differential}) can be integrated analytically for
arbitrary $\widetilde T_e$ to find $t
(\widetilde T_e)$.  If we neglect the pre-shock electron temperature as
well as energy exchange within the shock, and assume $\widetilde
T_{e2} = 0$ at the time $t_s$
when a particular fluid element has just passed through the shock, then
\begin{equation}
\Delta t \defeq t-t_s = t_{2s} \left ({n_i \over n_i + n_e} \right )
\left [\ln \left ({1 + \sqrt{ \widetilde T_e} \over 1-\sqrt{ \widetilde T_e}}
\right ) -2 \sqrt{ \widetilde T_e} \left (1 + {\widetilde T_e \over 3} \right )
\right ].
\label{solution}
\end{equation}
This solution has precisely the same form found by Shafranov (1957) in the
case of a stationary planar shock without post-shock adiabatic compression.
A nonzero value of $\widetilde T_{e2}$ would be equivalent to adding a
correction, $\delta t$, to $\Delta t$.  Because of the rapid relaxation
when $\widetilde T_e \ll 1$, this correction is generally $\ll t_{2s}$.
Thus at late times, when $\widetilde T_e \gg \widetilde T_{e2}$, the
corresponding correction to $\widetilde T_e (t)$ is small.  We will
estimate the magnitude of this effect in \S 5.

\subsection{Cluster Profile at a Given Time}

We now have an implicit solution for the scaled electron temperature,
$\widetilde T_e$, of a fluid element as a function of the time since that
fluid element passed through the shock.  For application to X-ray clusters,
one is more interested in $\widetilde T_e$ as a function of radius at a
fixed cosmic time, $t$.  We consider a fluid element observed at radius $r$ and
ask at what time, $t_s$, it passed through the shock.  The shock is
characterized by a fixed value, $\lambda_s$, of $\lambda = r/r_{ta}$.  For
a given fluid element,
\begin{equation}
{d \lambda \over d \ln t} = t { \dot r \over r_{ta}}
-t {r \over r_{ta}} {\dot r_{ta} \over r_{ta}}.
\label{eq:18}
\end{equation}
Expressing $\dot r = v$ in terms of the dimensionless $V = vt/r_{ta}$
and noting that $t \dot r_{ta}/r_{ta}= 8/9$ for infall onto a
top-hat perturbation, we can rewrite equation~(\ref{eq:18}) as
\begin{equation}
{d \lambda \over d \ln t} = V -{8 \over 9} \lambda ,
\end{equation}
so that
\begin{equation}
{t\over t_s} = \exp \int_\lambda^{\lambda_s} {d \lambda \over{8 \over 9}
\lambda -V} \defeq \epsilon (\lambda).
\label{define_epsilon} 
\end{equation}
The function $\epsilon(\lambda)$ is $1$ for $\lambda = \lambda_s$ and increases as $\lambda$
decreases.  For a fixed $t$, fluid elements at different $r$, and hence
different $\lambda$, have different values of $t_s$.  We are actually
interested in
\begin{equation}
\Delta t = t-t_s = t_s (\epsilon -1) = t (\epsilon - 1)/\epsilon.
\end{equation}

The timescale for relaxation is fixed by $t_{2s}$, which is a constant for
each fluid element but varies from one fluid element to the next.
Since $t_{2s} \propto \overline T_{2}^{3/2} n_{i2}^{-1}(t_s)$, we see from
the scaling of $\rho$ and $p$ at fixed $\lambda$
(eq.~[\ref{scalings}]) that
\begin{equation}
t_{2s} \propto \left. \left ({r_{ta} \over t} \right )^3 t^2 \right |_{t = t_s}
\propto t_s^{5/3} \propto \epsilon^{-5/3}.
\label{spatial_scaling}
\end{equation}

Following equation~(\ref{solution}), the scaled temperature, $\widetilde T_e$, is determined by the ratio
\begin{equation}
{\Delta t \over t_{2s}} =
{t \over t_{2}}
\epsilon^{ 2/3} (\epsilon -1).
\label{key_ratio}
\end{equation}
where $t_2$ is the value of $t_{2s}$ for the fluid element passing through
the shock at time $t$.  Combining this with the implicit solution for
$\widetilde T_e (\Delta t)$, and equation~(\ref{define_epsilon}), we have
an implicit solution for $\widetilde T_e$ as a function of $\lambda <
\lambda_s$.  Given $n_e/n_i$, $\widetilde T_e (\lambda)$ is
parameterized by the value of the ratio 
$t/t_{2}$.  These solutions are shown in Figure~1 for various
values of this parameter, using $n_e/n_i = 1.07$ as appropriate for a
fully ionized hydrogen-helium plasma with $X = 76 \%$.
For $t/t_{2} \gg 1$, thermal equilibrium takes place rapidly compared to
the dynamical time, and the temperature difference is significant only
in a narrow region near $r_{\rm shock}$, where the fluid elements have
passed through the shock very recently.  When $t/t_2 \la 1$, the effect
may be significant over a larger fraction of the shock radius.  The
detailed shape of the curves depends on $\epsilon (\lambda)$, which is
determined by the velocity of the fluid elements as they settle toward
their equilibrium radii. 

\section{Model Parameters}

To determine the significance of the lack of electron-ion equilibrium, we
need to relate the model parameters to values for typical observed
clusters.  The self-similar model itself has only one free parameter,
$r_{ta}$, which determines its scale.  The mass, density, and temperature
all depend on $r_{ta}$.  Unfortunately, X-ray measurements of these
quantities tend to come from the inner parts of clusters, where the
approximations leading to the self-similar solution break down and the
solution has unrealistic divergences.  We need to relate an observable
quantity in the inner regions of a cluster to a quantity in the outer
regions where self-similarity holds.  One way to do this is to use a
simplified model, such as that considered by Eke, Cole \& Frenk (1996), in
which both the X-ray gas and total mass have the form of a singular
isothermal sphere in the virialized section of the cluster.  In this model,
the mass within a mean overdensity of $178$ and the temperature of the
X-ray gas are related by
\begin{equation}
m_{178} =  2.34 \times 10^{15} h_{50}^{-1} \MSun 
\left [\left ({\Tx \over \temperature} \right )
{1 \over (1 + z)} \right]^{3/2},
\end{equation}
where $z$ is the cosmological redshift, and where we have assumed $X=76
\%$.  While this model is obviously an oversimplification, the relationship
it predicts between mass and emission-weighted X-ray temperature agrees
well with N-body/hydrodynamic simulations (Navarro et al. 1995; Pen 1997).
Furthermore, it meets our need to relate an observable quantity to a
property of the outer parts of clusters, where self-similarity might hold.

It is easy to show that the overdensity in Bertschinger's collisionless
model, given by the dimensionless quantity $M (\lambda)/\lambda^3$,
equals $178$ at $\lambda \approx 0.255$ where $M \approx 2.98$.  Using
the scaling relation equation~(\ref{mass_scaling}), we find
\begin{eqnarray}
r_{ta}  &=& 13.2 \, h_{50}^{-1} \Mpc \, (1 + z)^{-1} \left ({m_{178} \over \mass}
\right )^{ 1/3} \nonumber \\
 & = & 14.0  \, h_{50}^{-1} \Mpc \, (1 + z)^{-3/2}
\left ({\Tx \over \temperature}\right ) ^{ 1 / 2}.
\end{eqnarray}
The shock occurs at a fraction $\lambda_s = 0.3472$ of the turnaround
radius, so
\begin{eqnarray}
r_s  &=& 4.60 \, h_{50}^{-1} \Mpc \, (1 + z)^{-1} \left ({m_{178} \over \mass}
\right )^{ 1/3} \nonumber \\
 & = & 4.84 \, h_{50}^{-1} \Mpc \, (1 + z)^{-3/2}
\left ({\Tx \over \temperature}\right ) ^{ 1 / 2}.
\end{eqnarray}
Given $r_{ta}$ and $t = t_0 (1 + z)^{-3/2}$, we can convert the
dimensionless quantities $D$, $P$, and $V$ of the self-similar solution to
the physical quantities $\rho$, $p$, and $v$.  In the self-similar model,
the temperature $T \prop P/D$, which diverges as $\lambda^{-1/4}$ at small
$\lambda$.  Since we do not expect the self-similar model to be realistic
at small radii, we simply cut it off at $\lambda = 0.105$ where $T$ becomes
greater than $\Tx$.
For typical cluster conditions, the electrons and ions reach thermal
equilibrium outside this radius, and our results depend only on the
calibration of $r_{ta}$, not on the details of the central temperature
profile.

In order to calculate the equilibration timescale, $t_2$, we need the mean
temperature, $\overline T_2$, the ion density, $n_{i2}$, and the Coulomb
logarithm, $\Lambda_2$, all evaluated just inside the shock radius.  Given
the mean molecular weight, $\mu m_p$ (which we take to be $\mu = 0.59$ for
$X=76\%$), we can obtain $\overline T_2$ from $p_2/\rho_2$.  Using
Bertschinger's values for $P_2$ and $D_2$, we find
\begin{equation}
\overline T_2 = 0.44 \, \Tx.
\end{equation}
Similarly, we can find $n_{i2}$ and $n_{e2}$ from $\rho_2$.
\begin{equation}
\left \{{n_{i2} \atop n_{e2}} \right \}
=
\left \{{3.74 \atop 4.01} \right \}
\times 10^{-6} \percc 
\left ({\Omega_b h_{50}^2 \over 0.10} \right ) (1 + z)^3.
\label{densities}
\end{equation}
The Coulomb logarithm, which depends on $\overline T$ and $n_e$, is then
\begin{equation}
\ln \Lambda_2 = 39.7 + \ln\left({\Tx\over\temperature}\right)
 -\shalf \ln \left({\Omega_b h_{50}^2 \over 0.10} \right )
-\sthreehalves \ln (1 + z).
\end{equation}
Combining these values, we find 
\begin{equation}
{t \over t_2} = 0.828 \, h_{50}^{-1}
(1+z)^{3/2} \left ({\Tx \over \temperature} \right )^{-3/2}
\left ({\Omega_b h_{50}^2 \over 0.10} \right )
\left ({\ln \Lambda_2 \over 39.7} \right )
\end{equation}
Measurements of deuterium abundance in quasar absorption systems indicate
that $\Omega_b h_{50}^2 = 0.096^{+0.024}_{-0.020}$ (Tytler, Fan, \& Burles
1996; Burles \& Tytler 1997).  We assume $\Omega_b h_{50}^2 = 0.10$.  

\section{Numerical Calculation}

In order to produce cluster profiles at a fixed time, we need to reproduce
the self-similar solutions for $D$, $P$, and $V$.  We integrated the fluid
equations for one-dimensional self-similar accretion in an external
potential, following the same techniques as Bertschinger (1985).
Specifically, we integrated Bertschinger's equation (3.4a--c), with $M$
replaced with the collisionless $M_x$ appropriate for the $\Omega_b \ll 1$
case, using the change of variables given by Bertschinger's equation (3.12)
to avoid singularities at the origin.  The initial post-shock conditions
for the integration were given by Bertschinger's equation (3.6), where
$V_1$ and $D_1$, as well as the location, $\lambda_s = 0.3472$, of the
shock, were taken from Bertschinger's Table~10.  For simplicity, we
calculated the collisionless mass distribution, $M_x$, by quadratic
interpolation of Bertschinger's Table~4, using the scaling $M \propto
\lambda^{3/4}$ for $\lambda < 0.02$.

The equations were numerically integrated inward from the shock using a
fifth-order Runge-Kutta routine.  Values of $D$, $P$, and $V$ were
calculated on a grid with $\Delta \lambda = \lambda/500$.
The shock location was determined by the condition that the velocity,
$V$, should be zero at the origin.  Because we use values for $M_x$
interpolated from a table, we are effectively using a slightly different
mass distribution for the collisionless fluid.  Thus $\lambda_s = 0.3472$ is
no longer precisely the right value.  This causes discrepancies with
Bertschinger's $\Omega_b \ll 1$ solution which are worst near the
origin.  
Since we cut off our solutions for the electron and ion temperatures at
$\lambda = 0.105$ where $T = \Tx$, only the region outside that radius
concerns us.  In this region, $P/D$, which determines $\overline T$,
differs from Bertschinger's results by less than $0.5 \%$.  While the
fractional errors in $V$ reach $7 \%$, the resulting fractional errors in
$\epsilon$ are only $0.5 \%$.

We also used the two local integrals of motion given in Bertschinger's
equations (2.27) and (3.10) as checks on the calculation.  For $\lambda \ge
0.105$, the largest fractional errors in these integrals were $10^{-4}$.

In Figure~2, we show the electron, ion, and mean temperatures as a function
of radius for various values of the emission-weighted X-ray temperature.
In making comparisons with observed X-ray clusters, the density could in
principle be measured directly, rather than calculated from equation
(\ref{densities}).

\section{Approximations}

In \S 2, we made a number of approximations which we now proceed to
justify given the model parameters estimated in \S 3.  Several of the
approximations result in errors near the shock, where the temperature
equilibration is rapid due to the small value of $\widetilde T_e$.
The effect of these errors at later times is equivalent to a change of
definition of $\Delta t$ from $t-t_s$ to $t-t_s-\delta t$.
Because the relaxation rate for $\widetilde T_e \sim 1$ is much smaller
than when $\widetilde T_e \ll 1$, the  error
$\delta \widetilde T_e \approx \delta t (d \widetilde T_e/dt)$
at later times  is generally quite small.  One such approximation was
the assumption that $\widetilde T_e = 0$ at $ t_s$, neglecting the
pre-shock electron temperature  
as well as energy exchange within the shock.
In this case, $\delta t$ is the time it would take for $\widetilde T_e$
to go from zero to the actual $\widetilde T_{e2}$, which equation
(\ref{solution}) shows is
\begin{equation}
\delta t = t_{2s} \left ({n_i \over n_i + n_e} \right )
\left [\ln \left ({1 + \sqrt{ \widetilde T_{e2}} \over 1-\sqrt{
\widetilde T_{e2}}}
\right ) -2 \sqrt{ \widetilde T_{e2}} \left (1 + {\widetilde T_{e2} \over 3} \right )
\right ].
\label{t_correction}
\end{equation}
Thus, the fractional error in $\widetilde T_e$ is 
\begin{eqnarray}
{\delta \widetilde T_e \over \widetilde T_e} &=&
\left [\ln \left ({1 + \sqrt{ \widetilde T_{e2}} \over 1-\sqrt{
\widetilde T_{e2}}}
\right ) -2 \sqrt{ \widetilde T_{e2}} \left (1 + {\widetilde T_{e2}
\over 3} \right )
\right ]
(1- \widetilde T_e) \widetilde T_{e}^{-5/2} \cr 
& \approx & 0.4 \, (1- \widetilde T_e) \left ({\widetilde
T_{e2} \over \widetilde T_e} \right )^{5/2}
\end{eqnarray}
for $\widetilde T_{e2} \la 0.1$.  This correction is
small except where $\widetilde T_e$ is of order $\widetilde T_{e2}$.

To determine the significance of this effect, we need to know $\widetilde
T_{e2}$.  For $\overline T_2 = 0.44 \, \Tx$, adiabatic compression alone
results in $\widetilde T_{e2} = 5.7 \, T_{e1}/\Tx$.  Thus, in X-ray
clusters with $\Tx \ga 10^7 \K$, the resulting $\widetilde T_{e2}$ will be
small unless the pre-shock temperature is $\ga 10^6 \K$.

The contribution of collisional energy exchange within the shock to
$\widetilde T_{e2}$ depends on the detailed shock structure, but we can
estimate it.  If we repeat the derivation of \S 2.3, but ignore adiabatic
compression and scale the temperatures in terms of $\overline T_2$ instead
of $\overline T$, the factor of $(1-\widetilde T_e)$ in equation
(\ref{differential}) is replaced with $\overline T/\overline T_2-\widetilde
T_e$.  This reduces $d \widetilde T_e/dt$ by a factor $\ge \overline
T_2/\overline T$.  The resulting $\widetilde T_{e2}$ is given by equation
(\ref{t_correction}), where now $\delta t$ is the time spent in the shock
multiplied by the time-average of $\overline T/\overline T_2$.  As a rough
estimate, suppose this effective time, $\delta t$, equals the post-shock
proton collision time, $t_{pp2}$, which results in $\widetilde T_{e2} =
0.66$.  Figure 3 compares the resulting electron temperature profile with
the $\delta t = 0$ result for $\Tx = 10^8 \K$.  The fractional difference
is large at the shock, 
but quickly becomes smaller than the difference
between $T_e$ and $\overline T$.  As a fraction of $(1-\widetilde T_e)$,
the effect of energy exchange within the shock is less dramatic for lower
temperature clusters.  We should note that for $\delta t = t_{pp2}$, the
width of the shock is of order $1 \Mpc \times (\Tx/\temperature)^2$ at $z =
0$, which in the hottest clusters is a significant fraction of the shock
radius.  This will break self-similarity and alter the overall density and
temperature profiles of the cluster.  Figure 3 should only be regarded as
an approximate description for the electron temperature in this case.
Conversely, as long as a thin shock is a reasonable approximation (e.g. due
to embedded magnetic fields), collisional energy exchange within the shock
will not significantly alter the results for the electron temperature over
most of the region where $T_e$ and $T_i$ differ.

A second approximation which may fail near the shock is the assumption that
$T_e/T_i \gg m_e/m_p$.  For $\widetilde T_e \la 10^{-3}$, this breaks down,
so the equilibration timescale levels off, rather than continuing to
decrease as $\widetilde T_e^{3/2}$, and $\widetilde T_e$ grows only
exponentially, rather than as $\propto (\Delta t)^{2/5}$.  If $\widetilde
T_{e2}$ is less than this value, our assumption results in a shift, $\delta
t$, opposite in sign to the one above.  Given $\widetilde T_{e2}$ estimated
before, this is likely to be unimportant.  
Even for $\widetilde T_e \la 10^{-3}$, 
$t_{ei} \sim 10^{-4} t_{2s}$, so the time $| \delta t |$ to go
from $T_e \sim 10^3 \K$ to $\widetilde T_e= 10^{-3}$ is less than $4 \times
10^{-4} t_{2s}$. 
Thus, even in this worst case, $\delta\widetilde T_e$
remains significant only as long as $\widetilde T_e \la 10^{-3}$.

A final approximation which causes errors near the shock is our use of
$\widetilde T_e = 1$ in $\ln \Lambda_{2s}$.  The error introduced depends
on the value of $\ln \Lambda_{2s}$, formally breaking the one-parameter
family of solutions shown in Figure~1.  However, for the values of $\ln
\Lambda_{2s}$ calculated in \S 3 and shown in Figure~2, we can estimate its
magnitude.  A rigorous upper bound on this effect can be calculated, and
for $T_{e2} \ge 1000 \K$, it produces $\delta
\widetilde T_e/\widetilde T_e < 5 \%$ for $\widetilde T_e > 0.1$.

Two other approximations also involve $\ln \Lambda$,  but are exact at
the shock and become less accurate at smaller radii.  These effects
also break the one-parameter family of solutions and affect both
Figure~1 and Figure~2.  First, our scaling for  $t_{2s}$ for a
particular fluid element, given in equation (\ref{timescale_scaling}),
ignores the variation of $\Lambda$ with $\overline T$ and $n$ as the
fluid element is adiabatically compressed.  Since $\Lambda \sim
\overline T/\sqrt{ n} \sim \overline T^{1/4}$ for adiabatic
compression, and since $\overline T$ changes only by a factor of $2.2$
from the shock to the centrally-dominated $\Tx$, the fractional effect
on $\ln \Lambda$ is less than $0.6 \%$ and the integrated effect is
even smaller.  Second, the scaling used in
equations~(\ref{spatial_scaling}) and (\ref{key_ratio}) to relate
$t_{2s}$ to $t_2$ ignores the variation in $\ln \Lambda_{2s}$ between
fluid elements, since it replaces $\ln \Lambda_{2s}$ with $\ln
\Lambda_2$.  However, for the values of  $\Tx$ used in Figure 2, the
maximum fractional error in $\widetilde T_e$ is $0.1 \%$.

\section{Discussion}

Our model predicts significant differences between the electron and ion
temperatures in regions extending up to a third of the way inward from
the accretion shock.  The effect is greatest in the hottest clusters,
where the equilibration time, $t_{ei}$, is longest.  Because the
equilibration is exponential (except at early times)  with a timescale
neither much longer nor much shorter than the ages of
clusters, it is difficult to predict the extent of nonequilibrium
precisely.  As more direct information about the physical characteristics
of the X-ray gas in the outer parts of clusters becomes available, it
will be possible to refine these predictions.

In principle, the ion temperature
could be inferred by measuring the width of X-ray lines broadened by the
Doppler shift due to the ion thermal velocity dispersion.  The
equilibration timescale, $\equilibrium$, is proportional to $A/Z^2$, so
the high ionization states which dominate at cluster temperatures should
equilibrate with the protons on timescales $\le t_{pp}$.  Thus we can
assume a single ion temperature, $T_i$.  The rms velocity dispersion
due to thermal motions along the line of sight is $\los = (k
T_i/m_i)^{ 1 / 2}$, which produces a line with an rms width
\begin{equation}
{\Delta E_{\rm rms} \over E} = {\los \over c}
= {1 \over c} \sqrt{ k T_i \over m_i}
= {3.0 \times 10^{-3} \over \sqrt{ A} } \left ({T_i \over \temperature}
\right )^{ 1 / 2}
\label{resolving}
\end{equation}
where $A = m_i/m_p$.  In terms of a resolving power, $E/\Delta E$, where
$\Delta E$ is the {\it Full Width at Half Maximum} of a Maxwellian
velocity distribution, equation (\ref{resolving})
corresponds to
\begin{equation}
{E \over \Delta E} = 1.4 \times 10^{2} \sqrt{ A}
\left ({\temperature \over T_i} \right )^{ 1 / 2}.
\end{equation}
For the iron lines near $7 \keV$, a resolving power of $E/\Delta E
\approx 1000$, or $\Delta E = 7 \eV$, would be needed.  The
microcalorimeters of the XRS instrument on the planned Astro~E satellite
have an instrumental $\Delta E \approx 11.5 \eV$.  Convolution with the
intrinsic width would broaden this to an observed $\Delta E$ of $13.5
\eV$.  A high signal-to-noise ratio, as well as an accurate calibration
of the instrumental profile, would be necessary to detect this
broadening.

The HETG spectrometer on AXAF will reach the necessary resolving power of
500 -- 1000 at energies $\la 2 \keV$, for observations of point sources.
It might therefore be able to measure the widths of absorption lines of
\ion{O}{8}, \ion{Ne}{10}, or \ion{Mg}{12} in the spectra of quasars located
behind the cluster.  While the ionization state of these elements is
determined by $T_e$ rather than $T_i$, the populations of these states are
still likely to be small in the hotter clusters where we expect significant
differences between $T_e$ and $T_i$.

Turbulent line broadening may also be significant in the outer parts of
clusters.  The ratio of thermal to turbulent broadening is
\begin{equation}
{\los\over \turbulent} = \left({\sound\over \turbulent}\right) \sqrt{k
T_i/A m_p \over \gamma k T/\mu m_p} \approx {\sound/\turbulent \over 1.7
\sqrt{A}}
\end{equation}
where $\sound$ is the sound speed.  Widths of several lines of
species with
$A \la (\sound/\turbulent)^2$ would be necessary to distinguish thermal
broadening from turbulent broadening.

Fine structure lines in the ultraviolet might provide another probe of the
ion temperature.  At the temperatures in question, proton impact excitation
can contribute significantly to the population of the upper levels of the
\ion{Fe}{18} $\lambda974$ and \ion{Fe}{21} $\lambda 1354$ lines (Raymond
1997). Its effect on these lines could therefore be used to measure the 
ion temperature.

Differences between electron and ion temperatures might be even more
significant in clusters formed by mergers than in our simple, spherical
accretion model.  In a merger, the time since gas was shocked would be
smaller, but preheating of the gas would be more important.
Three-dimensional hydrodynamic simulations would be essential in exploring
the extent of deviations from thermal equilibrium in this case.

\acknowledgements

We thank Bill Forman, Maxim Markevitch, Ue-Li Pen, and John Raymond for
useful discussions. This work was supported in part by a National Science
Foundation Graduate Research Fellowship and by NASA GSRP Fellowship
NGT-51664 for DCF, and by NASA ATP grant NAG5-3085 and the Harvard
Milton fund for AL.  One of the authors (DCF) would like to acknowledge
Dragon Systems, Inc., whose DragonDictate for Windows software was
essential to the preparation of this paper.


\clearpage
\newpage
\begin{figure}[b]
\vspace{2.6cm}
\includegraphics{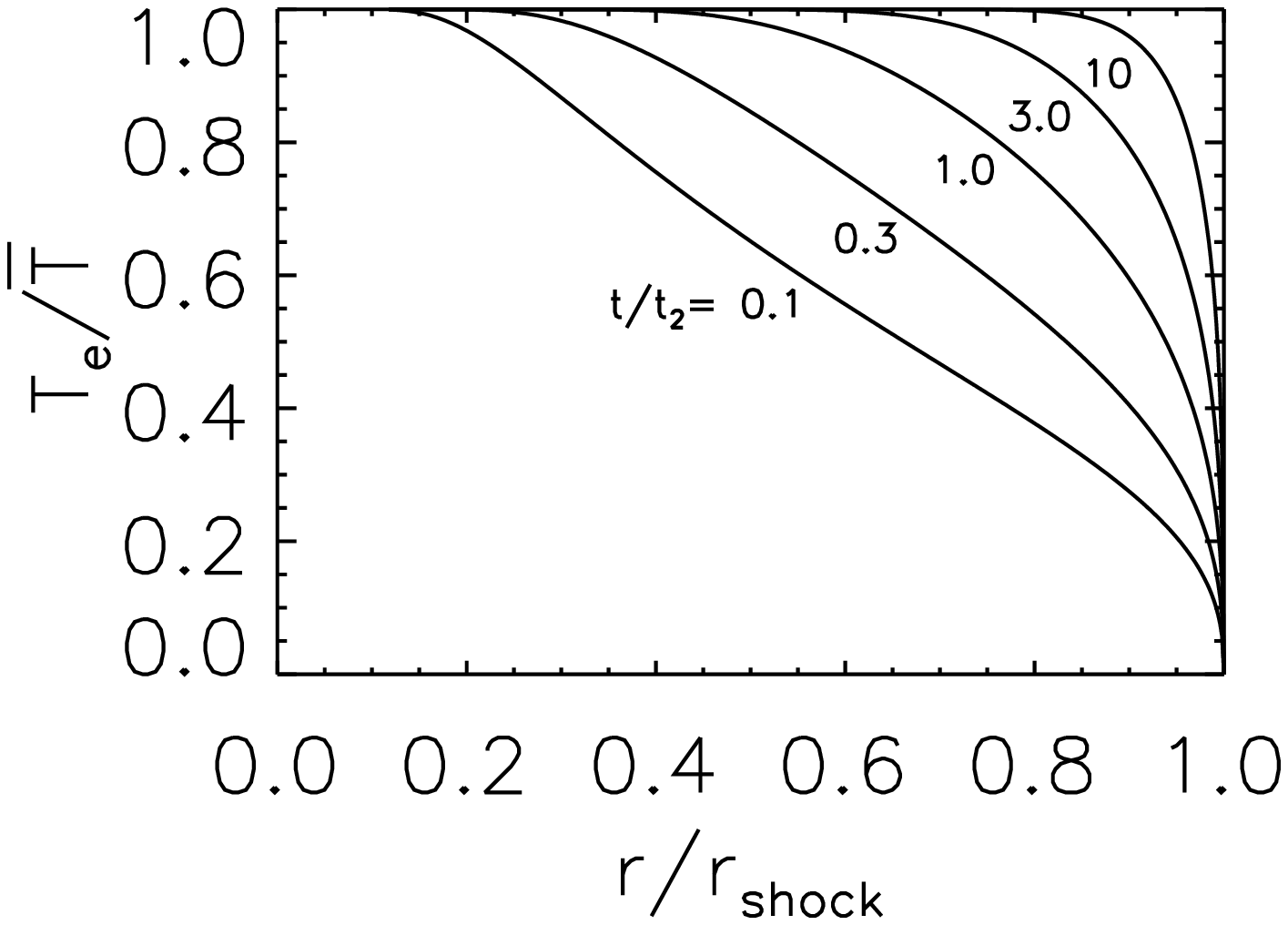}
\vspace*{4.5in}
\caption[scaled] {Scaled electron temperature,
 $T_e/\overline T$, as a function of scaled radius, $r/r_{\rm shock}$,
for various values of the ratio between
the cosmic time $t = t_0/(1 + z)$ and the
electron-ion equilibration time at the shock, $t_2$.}
\end{figure}

\clearpage
\newpage
\begin{figure}[b]
\vspace{2.6cm}
\includegraphics{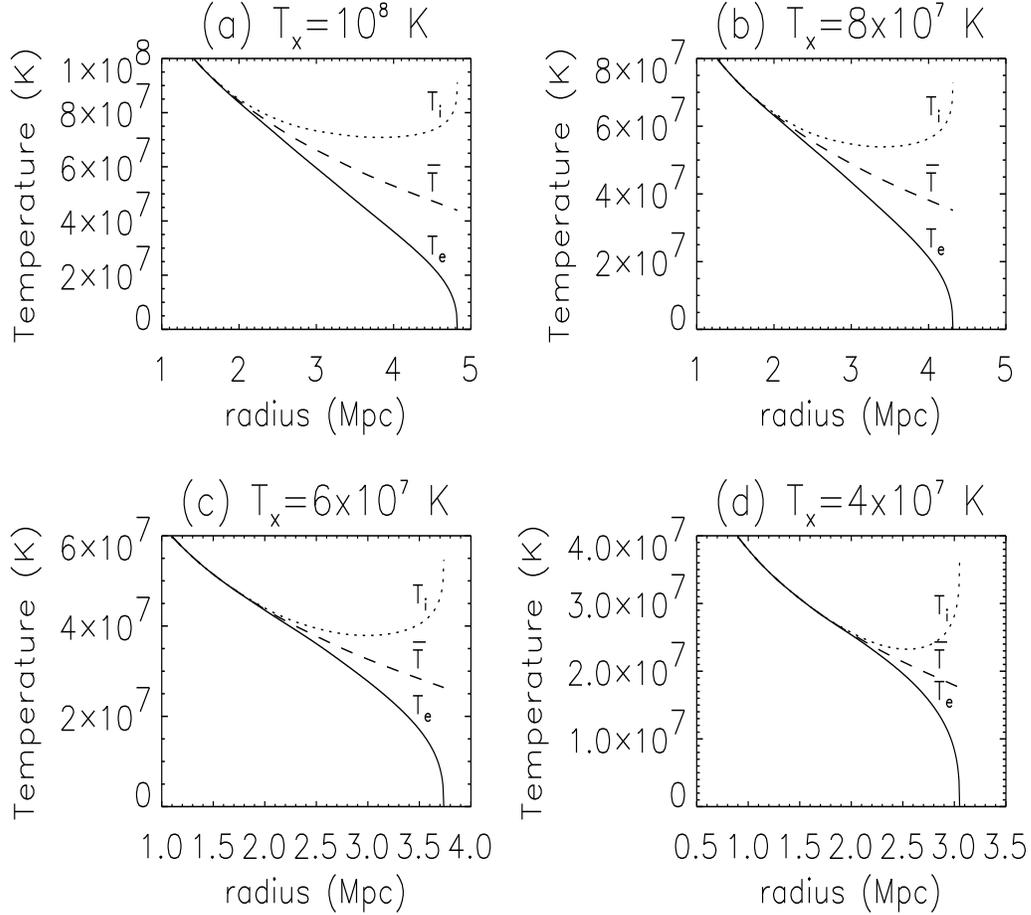}
\vspace*{4.5in}
\caption[realistic] {Electron temperature (solid), ion temperature (dotted), 
and mean temperature (dashed) as a function of radius for four 
clusters with emission-weighted
X-ray temperatures, $\Tx$, of (a) $10^8 \K$, (b) $8 \times 10^{7} \K$, (c)
$6 \times 10^{7} \K$, and (d) $4 \times 10^{7}
\K$.}
\end{figure}

\clearpage
\newpage
\begin{figure}[b]
\vspace{2.6cm}
\includegraphics{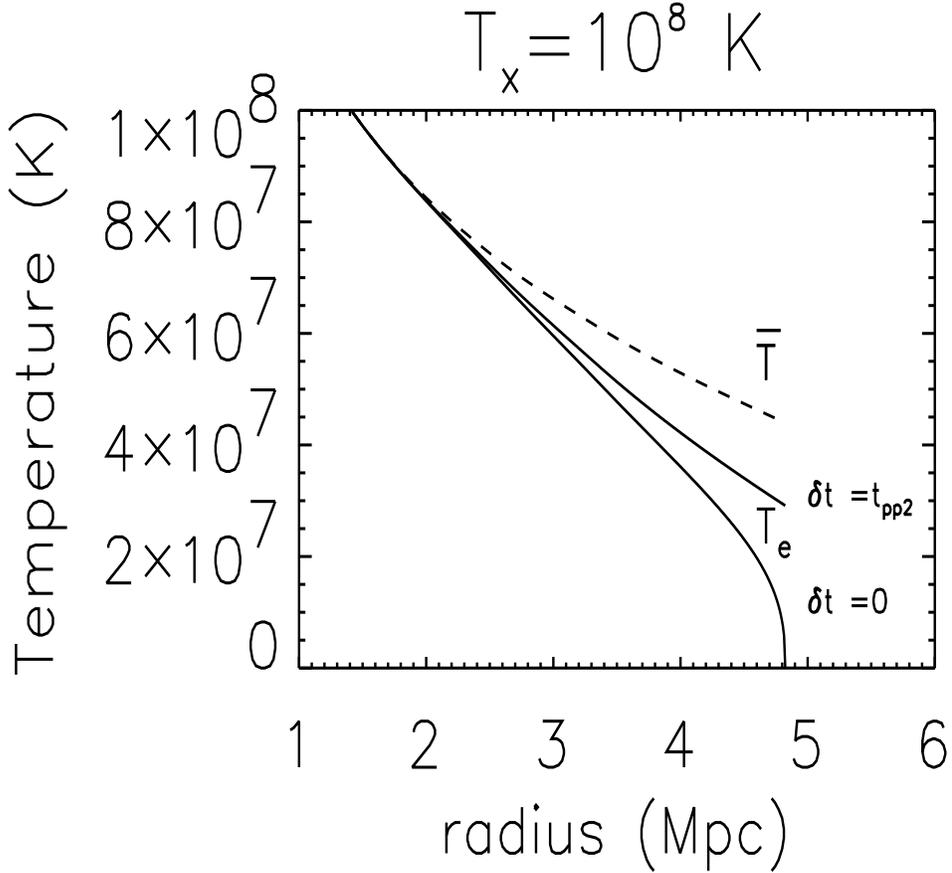}
\vspace*{4.5in}
\caption[exchange] {Effect of energy exchange within the
shock for a cluster with emission-weighted temperature $\Tx = 10^{8}
\K$.  The electron temperature (solid) 
assuming an effective time within the shock of one post-shock
proton-proton collision time ($\delta t = t_{pp2}$) is compared with the case
of no exchange ($\delta t = 0$).  The mean temperature (dashed) is also
shown.  The effect is smaller for lower
temperature clusters.}
\end{figure}

\end{document}